\documentclass[groupedaddress,
  aps,
  prl,
  amsmath,amssymb,
 reprint,
]{revtex4-1}
\usepackage{graphicx,color}
\usepackage{amsmath}
\usepackage{natbib}
\usepackage{epsfig}
\begin{document}

\title{Lindemann melting criterion in two dimensions}

\author{Sergey A. Khrapak}
\email{Sergey.Khrapak@dlr.de}
\affiliation{Institut f\"ur Materialphysik im Weltraum, Deutsches Zentrum f\"ur Luft- und Raumfahrt (DLR), 82234 We{\ss}ling, Germany;\\ Joint Institute for High Temperatures, Russian Academy of Sciences, 125412 Moscow, Russia}

\date{\today}

\begin{abstract}
It is demonstrated that the Lindemann's criterion of melting can be formulated for two-dimensional classical solids using statistical mechanics arguments. With this formulation the expressions for the melting temperature are equivalent in three and two dimensions. Moreover, in two dimensions the Lindemann's melting criterion essentially coincides with the Berezinskii-Kosterlitz-Thouless-Halperin-Nelson-Young melting condition of dislocation unbinding.       
\end{abstract}

\maketitle

The famous Lindemann's melting criterion~\cite{Lindemann} states that melting of a three-dimensional (3D) solid occurs when the square root of the particle mean squared displacement (MSD) from the equilibrium position reaches a threshold value (roughly $\sim 0.1$ of the interparticle distance). This is the oldest and apparently the most widely used method to {\it approximately} predict melting parameters of real and model systems. The conventional Lindemann's criterion is not applicable to the two-dimensional (2D) solid, because long-wavelength density fluctuations cause the MSD to diverge logarithmically with system size~\cite{Peierls1934,Landau1937,IllingPNAS2017}. This divergence implies the absence of long-range order and that this 2D ``solid'' is not a solid in the usual sense. Nevertheless, numerous examples of (finite) 2D crystals exist, ranging from atomic monolayers and thin films on a substrate ~\cite{Kosterlitz1978}, electron layers on the surface of liquid helium~\cite{GrimesPRL1979}, to colloidal particles at flat interfaces~\cite{ZahnPRL1999,ZahnPRL2000,KelleherPRE2017} and complex (dusty) plasmas in ground-based conditions~\cite{ChuPRL1994,ThomasPRL1994,MelzerPLA1994,Hayashi1994,ThomasNature1996,FortovPR,MorfillRMP2009}.     

The apparent controversy between the absence of long-range order in 2D and computer simulation results, evidencing the existence of 2D crystals, stimulated investigations into the nature of the fluid-solid phase transition in 2D systems and its difference from the 3D scenario. As a results, the celebrated Berezinskii-Kosterlitz-Thouless-Halperin-Nelson-Young (BKTHNY) scenario emerged~\cite{KosterlitzRMP2017,BerezinskiiJETP1971,KosterlitzJPC1973,NelsonPRB1979,YoungPRB1979,RyzhovUFN2017}. According to the BKTHNY theory, melting is a two-stage process. The crystal first melts by dislocation unbinding to an anisotropic hexatic fluid and then undergoes a continuous transition into an isotropic fluid. This scenario has been confirmed by numerical simulations~\cite{LinPRE2006} and colloidal experiments~\cite{ZahnPRL1999,ZahnPRL2000,DeutschlanderPRL2014,KelleherPRE2015}.  It is also understood that the 2D melting scenario depends considerably on the potential softness~\cite{KapferPRL2015}. BKTHNY scenario operates in systems with sufficiently soft long-range interactions~\cite{KapferPRL2015}. On the other hand, for steeply repulsive interactions the hard-disk melting scenario holds with a first-order liquid-hexatic and a continuous hexatic-solid transition~\cite{BernardPRL2011,EngelPRE2013,ThorneyworkPRL2017}.

Although MSD diverges with system size in 2D crystalline and amorphous solids~\cite{IllingPNAS2017,ShibaPRL2016}, the squared deviation of the difference between the positions of two particles remains finite~\cite{JancoviciPRL1967}. This can serve as a basis to construct modified Lindemann-like criteria of 2D melting. For example, the ratio of mean square difference of displacements in neighboring lattice cites to the square of the interparticle distance was proposed to serve as a modified Lindemann's criterion in 2D~\cite{LozovikSSC1985,BedanovPLA1985,GoldoniPRB1996}. Later, a related observation was reported that when measuring the displacements of particles in local coordinate systems, the Lindemann's criterion appears to apply also in 2D~\cite{ZhengEPL1998}. Following the same lines, a dynamical Lindemann-like measure has been introduced in Refs.~\cite{ZahnPRL1999,ZahnPRL2000}. Nevertheless, all such modifications remain {\it ad hoc}.
       
The purpose of this Letter is to demonstrate that the Lindemann's melting criterion can be reformulated for 2D classical systems using statistical mechanics arguments. This reformulation is similar in sense, but not identical, to that proposed by Ross for the 3D case~\cite{RossPR1969}. The conventional 3D Lindemann's melting rule and its modified 2D variant result in equivalent expressions for the melting temperature in 2D and 3D cases. Moreover, the expression for the 2D Lindemann's melting rule essentially coincides with the BKTHNY melting condition.  

It is useful to remind first the main steps relating the Lindemann's melting criterion and the low-frequency collective modes. For $N$ identical particles forming a crystalline solid the MSD is
\begin{equation}\label{MSD1}
\langle\xi^2\rangle =\frac{1}{N}\sum_i \langle \xi_i^2\rangle =\frac{1}{N}\sum_{{\bf k}}\langle \xi_{\bf k}^2 \rangle,
\end{equation} 
where as usually the summation over particles has been replaced by the summation over normal modes characterized by wavevectors ${\bf k}$. Mainly for the sake of simpler notation, high symmetry crystals are considered, so that quantities such as MSD or sound velocities can be considered as isotropic to a good approximation. From the energy equipartition we have
\begin{equation}\label{equipartition}
\frac{1}{2}m\omega_{\bf k}^2 \langle \xi_{\bf k}^2 \rangle =\frac{1}{2}T,
\end{equation}
where $m$ is the particle mass, $\omega_{\bf k}$ is the frequency associated with the wavenumber ${\bf k}$, and $T$ is the temperature in energy units ($k_{\rm B}=1$). This results in
\begin{equation}\label{MSD2}
\langle \xi^2 \rangle = \frac{T}{mN}\sum_{\bf k}\frac{1}{\omega_{\bf k}^2}=\frac{{\mathcal D}T}{m}\left\langle {\frac{1}{\omega^2}}\right\rangle,
\end{equation}
where ${\mathcal D}$ is the number of spatial dimensions and hence ${\mathcal D}N$ is the number of normal modes. The averaging can be performed using the vibrational density of states (VDOS) $g(\omega)$~\cite{YoungJCP1974}:
\begin{equation}\label{average}
\left\langle \frac{1}{\omega^2}\right\rangle=\int\frac{g(\omega)}{\omega^2}d\omega. 
\end{equation}
In the Debye approximation it is assumed that $g(\omega)\propto \omega^{{\mathcal D}-1}$ up to a cutoff frequency $\omega_{\rm D}$ and is zero otherwise. Combining this with the normalization condition $\int g(\omega)d\omega = {\mathcal D}N$, the MSD can be evaluated in 3D as $\langle \xi^2\rangle = 9T/m\omega_{\rm D}^2$. The Lindemann's criterion states that $\langle \xi^2 \rangle = L^2 a^2$, where $L$ is the Lindemann parameter and $a$ is the characteristic interparticle distance [in this paper $a$ is given by the corresponding Wigner-Seitz radius, that is $a=(4\pi n/3)^{-1/3}$ in 3D and $a=1/\sqrt{\pi n}$ in 2D]. The melting temperature can be roughly estimated from
\begin{equation}\label{Lindemann1}
T_{\rm m}\simeq Cm\omega_{\rm D}^2a^2,
\end{equation}
where $C$ is expected to be a quasi-universal constant.  

The Debye frequency $\omega_{\rm D}$ can be expressed in terms of longitudinal and transverse sound velocities and hence infinite frequency (instantaneous) elastic moduli. Taking into account that summation over normal modes can be replaced by integration $\sum_{\bf k}\rightarrow V\int d^3{\bf k}/(2\pi)^3$ and that the contribution from each polarization should ammount to $N$, we arrive at the normalization condition $(4\pi/3)(k_{\max}/2\pi)^3=n$ for each of polarizations. In 3D case we have one longitudinal and two transverse polarizations, characterized by the acoustic dispersion relations $\omega_{l}(k)=kc_l$ and $\omega_t(k)=kc_t$, where $c_l$ and $c_t$ are the longitudinal and transverse sound velocities, respectively. The sum of these contributions, each taken with its own cutoff $k_{\max}=\omega_{\rm D}/c_{l,t}$ should equal $3n$, which results in~\cite{BuchenauPRE2014}
\begin{equation}\label{wD1}
\omega_{\rm D}^3=18\pi^2 n\left(c_l^{-3}+2c_t^{-3}\right)^{-1}.
\end{equation}       
  
This consideration fails in 2D since the VDOS behaves as $g(\omega)\propto \omega$ and hence the integral (\ref{average}) diverges logarithmically in the thermodynamic limit. For finite systems the maximum wavelength that can be supported provides a lower limit for wave numbers $k_{\rm min}\sim 1/R$ and the corresponding lower limit for frequencies $\omega_{\rm min}\sim c_t/R$, where $R$ is the characteristic system size. Integration in Eq.~(\ref{average}) leads to the logarithmic divergence of the MSD with the system size: $\langle \xi^2\rangle \propto\ln(\omega_{\rm D}/\omega_{\rm min})\propto \tfrac{1}{2}\ln N$. This scaling has been repeatedly reproduced in molecular dynamics simulations of crystalline and amorphous solids~\cite{YoungJCP1974,GannPRB1979,ToxvaerdPRL1983,ShibaPRL2016,IllingPNAS2017}. This is the basis behind the conventional statement that the Lindemann's melting criterion does not exist in 2D dimensions.

However, as has already been mentioned, there exist alternative formulations. Apart from (to some extent {\it ad hoc}) definitions of the {\it local} Lindemann's criterion (either in terms of nearest neighbor displacements or local coordinates) a more physically justified approach exists, based on statistical mechanics arguments. It was Ross~\cite{RossPR1969} who proposed to generalize the conventional Lindemann's criterion in 3D in terms of the partition function. He argued that looking from the microscopic level on the melting transition, we should see the same scaled picture in the solid. For a given crystalline structure, the ratios of effective volumes occupied by atoms to the total volume of the system should remain constant along the melting curve. The relative atom arrangements is space should also remain the same. Consequently, the pictures along the melting curve should be identical if properly scaled, and this allows to express the Lindemann's melting law in terms of statistical mechanics~\cite{RossPR1969}. This point of view is further supported by the concept of isomorphs, which correspond to curves in the thermodynamic phase diagram along which structure and dynamics in properly reduced units are invariant to a good approximation~\cite{DyreJPCB2014,GnanJCP2009}. Melting and freezing curves appear as {\it approximate} isomorphs~\cite{PedersenNatCom2016}. Although the isomorphs concept is not yet well developed in 2D, here it is merely used to reinforce the original Ross's argumentation.   

How this argumentation applies to the melting of 2D solids? The starting point is the Helmholtz free energy of a 2D solid in the harmonic approximation~\cite{LL_StatPhys}
\begin{equation}
F=E_{\rm L}+T\int\ln\left[1-\exp\left(-\frac{\hslash \omega}{T}\right)\right]g(\omega)d\omega,
\end{equation}      
where $E_{\rm L}$ is the energy of all particles at their lattice cites (lattice sum), $\hslash$ is the Planck's constant, and the integration is from zero to the 2D Debye frequency. Taking the high-temperature limit $T\gg \hslash \omega$ and subtracting the free energy of an ideal 2D gas, $F_{\rm id}=-NT\ln\left[(e/n)(mT/2\pi\hslash^2)\right]$, the excess free energy becomes 
\begin{equation}\label{F_1}
F_{\rm ex} = E_{\rm L}+NT+NT\left\langle\ln\frac{m\omega^2 a^2}{2T}\right\rangle.
\end{equation} 
In his 3D derivation Ross further assumed the (Einstein) single-particle cell model, where each atom is confined within its cell and moves in a potential field of other stationary atoms located in the respective lattice cites~\cite{RossPR1969}. This would be an extreme oversimplification in 2D case. This becomes particularly evident by noting that within the Einstein model $\langle\xi^2 \rangle$ remains finite in 2D solids.

As a more convincing alternative, the averaging in Eq.~(\ref{F_1}) can be readily performed using the 2D Debye model with $g(\omega)\propto \omega$ and the result is
\begin{equation}\label{F2D}
F_{\rm ex}= E_{\rm L}+NT\ln\frac{m\omega_{\rm D}^2a^2}{2T}.
\end{equation}
The first term just depends on the amplitude of the interparticle interaction and is irrelevant in the present context. According to Ross's argumentation (or isomorphs concept) the second term should remain approximately constant. This immediately leads us to Eq.~(\ref{Lindemann1}) with a constant, which is potentially different from that in 3D.   

The 2D Debye frequency is found very similarly to the 3D one, but taking into account that the normalization condition is $\pi \left(k_{\rm max}/2\pi\right)^2=n$ for the longitudinal and transverse modes. The result is
\begin{equation}\label{wD2}
\omega_{\rm D}^2 = 8\pi n\left(c_l^{-2}+c_t^{-2}\right)^{-1}.
\end{equation} 
Taking into account the strong inequality $c_l^2\gg c_t^2$  (which holds in both 2D and 3D soft interacting particle systems~\cite{KhrapakSciRep2017,KhrapakMolPhys2019,KhrapakPoP2019})  we arrive at the 2D melting conditions
\begin{equation}\label{melt2D}
\frac{c_t^2}{v_{\rm T}^2}\left(1-\frac{c_t^2}{c_l^2}\right)\simeq {\rm const},
\end{equation}
where $v_{\rm T}=\sqrt{T/m}$ is the thermal velocity. This is the main result of this Letter, which will be scrutinized from several different perspectives below.  

The first obvious question is how the 2D Lindemann's melting criterion derived above is related to the established BKTHNY melting condition. In the BKTHNY theory of melting, the dislocation unbinding occurs when the Young's modulus reaches the universal value of $16\pi$,
\begin{equation}\label{KTHNY}
\frac{4\mu(\mu+\lambda)}{2\mu+\lambda}\frac{b^2}{T}= 16\pi,
\end{equation}
where $\mu$, $\lambda$ are the Lam{\' e} coefficients of the 2D solid, and $b$ is the lattice constant. The Lam{\' e} coefficients of an ideal 2D lattice can be expressed in terms of the sound velocities~\cite{PeetersPRA1987,LL_Elasticity} as $\mu = m n c_{t}^2$ and $\lambda=m n (c_{l}^2-2c_{t}^2)$. It is easy to show that the condition (\ref{KTHNY}) becomes identical to (\ref{melt2D}), provided the constant is fixed at ${\rm const} = 2\pi\sqrt{3}$. It is important to note, however, that the Lam{\' e} coefficients to be substituted in Eq.~(\ref{KTHNY}) should be evaluated taking into account (i) {\it thermal softening} and (ii) {\it renormalization} due to dislocation-induced softening of the crystal~\cite{MorfPRL1979,ZanghelliniJPCM2005}.  Original simplistic theoretical estimates using the elastic constants of an ideal crystalline lattice at $T=0$ yield melting temperatures overestimated by a factor between $\simeq 1.5$ and $\simeq 2$ for various 2D systems~\cite{ZanghelliniJPCM2005,ThoulessJPC1978,PeetersPRA1987,KhrapakCPP2016}. At the same time it has been demonstrated recently that a simple renormalization of the constant in Eq.~(\ref{KTHNY}) can approximately account for thermal and dislocation induced softening~\cite{KhrapakJCP2018_1}.  In this sense the 2D Lindemann's and KTHNY melting criteria can be viewed as essentially equivalent (at least for sufficiently soft interactions).     

The second natural question is how different are the constants in Eq.~(\ref{Lindemann1}) in the case of 3D and 2D geometries. To get some insight, let us consider the special case of repulsive Coulomb interaction potential $\varphi(r) = Q/r$, where $Q$ is electrical charge. This system is often referred to as the one-component plasma (OCP) and can be characterized by the single Coulomb coupling parameter $\Gamma=Q^2/aT$. In this special case, the long-ranged character of the potential makes the longitudinal dispersion non-acoustic, with $\omega\simeq \omega_{\rm p}$ in 3D (the plasma frequency in 3D is $\omega_{\rm p}=\sqrt{4\pi Q^2n/m}$) and $\omega\simeq \omega_{\rm p}\sqrt{ka}$ in 2D (the plasma frequency in 2D is $\omega_{\rm p}=\sqrt{2\pi Q^2n/ma}$ )~\cite{Baus1980,KhrapakJCP2018}. In the present context this simply implies $c_t/c_l=0$ and, hence, only the transverse sound velocity matters. The latter is proportional to the universal scaling factor $\sqrt{Q^2/\Delta m}$, where $\Delta=n^{-1/{\mathcal D}}$ is the interparticle separation. The proportionality constant is $\simeq 0.440$ in 3D and $0.495$ in 2D~\cite{KhrapakPoP2019}. The fluid-solid phase transition takes place at $\Gamma_{\rm m}\simeq 175$ in 3D~\cite{DubinRMP1999} and $\Gamma_{\rm m}\simeq 135$ in 2D~\cite{GrimesPRL1979}. This suffices to evaluate the involved constants. It turns out that the ratio $m\omega_{\rm D}^2a^2/T_{\rm m}$ is $\simeq 160$ in 3D and $\simeq 150$ in 2D. This corresponds to the Lindemann parameter $L\simeq 0.24$ both in 2D and 3D (note that here the Lindemann parameter is expressed in terms of the Wigner-Seitz radius $a$). 

It is not easy to give a general estimate regarding the suitability of the harmonic approximation. Nevertheless, it is instructive to compare the magnitudes of anharmonic terms for similar systems in 3D and 2D. Let us again consider the OCP model. The Helmholtz free energies of OCP solids have been summarized for instance in Ref.~\cite{KhrapakCPP2016}. It turns out that the leading (quadratic in temperature) anharmonic term is $\simeq 2$ times larger in 3D than in 2D. In relative units this term amounts to $\sim0.04\%$ (3D) and $\sim 0.02\%$ (2D) of the total free energy at the melting temperature (even such a small difference can, however, matter when looking for the intersection of fluid and solid free energy curves).

The next important observation is that, because of the strong inequality $c_l^2\gg c_t^2$, the melting indicator (\ref{melt2D}) can be simply reduced to the condition of constant transverse-to-thermal velocity ratio at melting:   
\begin{equation}\label{melting_tr}
\left. \frac{c_t}{v_{\rm T}}\right|_{T_{\rm m}}\simeq {\rm const}. 
\end{equation} 
The condition of this kind has been previously reached as a consequence of the BKTHNY melting condition~\cite{KhrapakJCP2018_1}. It now appears that this condition operates in both 3D and 2D geometries and can be regarded as a consequence of the generalized Lindemann's melting rule. The fact that the transverse sound velocity plays dominant role is yet another demonstration of the ``shear dominance'' effect~\cite{DyreRMP2006}. As previously, we can determine the constants in Eq.~(\ref{melting_tr}) by considering the Coulomb limit. This yields $c_t/v_{\rm T}\simeq 4.6 (4.3)$ at melting of a 3D (2D) solid. 

        
\begin{figure}
\includegraphics[width=7cm]{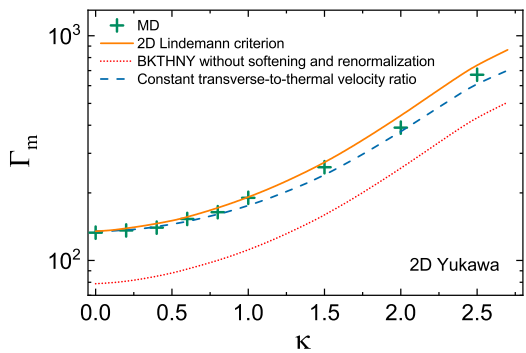}
\caption{Melting curve of a 2D Yukawa crystal in the ($\kappa$, $\Gamma$) plane. The symbols are the results of MD simulations~\cite{HartmannPRE2005}. The solid curve corresponds to the 2D Lindemann's melting rule of Eqs.~(\ref{Lindemann1}) and (\ref{wD2}). The dashed curve is plotted using Eq.~(\ref{melting_tr}). The dotted line corresponds to the solution of BKTHNY condition (\ref{KTHNY}) with the asymptotic $T=0$ values of elastic constants (i.e. without taking into account thermal softening and renormalization)~\cite{PeetersPRA1987}. }
\label{Fig1}
\end{figure}

As a demonstration of the 2D Lindemann's criterion at work, the melting line of the 2D Yukawa (Debye-H\"uckel) solid is shown in Figure~\ref{Fig1}. In Yukawa systems the particles interact via the pairwise repulsive exponentially screened Coulomb potential $\varphi(r)=Q^2\exp(-r/\lambda)/r$, where $\lambda$ is the screening length. The phase state is fully characterized by the two dimensionless parameters: the coupling parameter $\Gamma=Q^2/aT$ and the screened parameter $\kappa=a/\lambda$. The Yukawa interaction potential is often used as a first approximation to real interactions in systems of electrically charged particles, such as ions in aqueous solutions of electrolytes, colloidal suspensions, and complex (dusty) plasmas~\cite{IvlevBook,LowenPRL1993,FortovPR,ChaudhuriSM2011}.  
Phase diagrams of Yukawa systems have been extensively investigated both in 3D and 2D and are relatively well known~\cite{RobbinsJCP1988,MeijerJCP1991,HamaguchiPRE1997,VaulinaPRE2002,HartmannPRE2005,KhrapakPRL2009,YazdiPRE2014,
YurchenkoJPCM2016,KryuchkovJCP2017}. In Figure~\ref{Fig1} symbols correspond to MD simulation results from Ref.~\cite{HartmannPRE2005}, where the location of the melting line was determined from the analysis of the bond-angular order parameter. The solid curve corresponds to the 2D Lindemann's melting criterion of Eqs.~(\ref{Lindemann1}) and (\ref{wD2}). The sound velocities have been evaluated using the approach described in Refs.~\cite{KhrapakPoP2018,KhrapakPoP2019}. The dashed curve shows the application of a simplified melting indicator of Eq.~(\ref{melting_tr}). Both curves agree satisfactory with the results of MD simulation.   An early attempt to estimate the location of the melting curve by using Eq.~(\ref{KTHNY}) with the asymptotic $T=0$ values of elastic constants~\cite{PeetersPRA1987} is depicted by the dotted curve.
This curve is located considerably lower.    

\begin{figure}
\includegraphics[width=7cm]{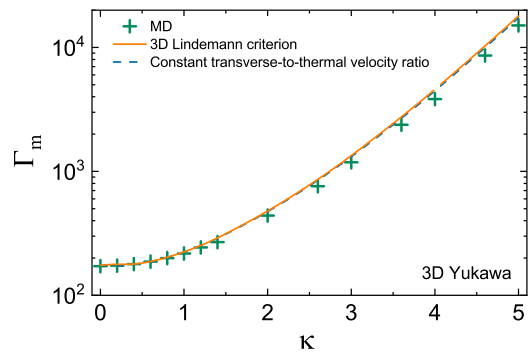}
\caption{Melting curve of a 3D Yukawa solid in the ($\kappa$, $\Gamma$) plane. Symbols are the results from MD simulations~\cite{HamaguchiPRE1997}. The solid curve corresponds to the 3D Lindemann's melting rule of Eqs.~(\ref{Lindemann1}) and (\ref{wD1}). The dashed line represents Eq.~(\ref{melting_tr}). }
\label{Fig2}
\end{figure}

To demonstrate simultaneous applicability of the Lindemann's law in both 2D and 3D, the melting line of a 3D Yukawa solid is plotted in Fig.~\ref{Fig2}. Symbols correspond to MD results from Ref.~\cite{HamaguchiPRE1997}, where the fluid-solid phase transition was identified from the free energy consideration. The solid curve corresponds to the 3D Lindemann's melting criterion of Eqs.~(\ref{Lindemann1}) and (\ref{wD1}), while the dashed curve corresponds to the simplified condition (\ref{melting_tr}). Both curves are in reasonable agreement with the MD results. Additionally we observe much smaller difference between the Lindemann's melting criterion (\ref{Lindemann1}) and its simplified version (\ref{melting_tr}) in 3D. This has the following explanation: First, the ratio of transverse-to longitudinal sound velocities is somewhat higher in 2D, and, second, there are two transverse modes in 3D, but only one in 2D. Both factors diminish the importance of the longitudinal sound in 3D case and the result of this can be clearly observed from comparing Figs.~\ref{Fig1} and \ref{Fig2}.            

\begin{figure}
\includegraphics[width=7cm]{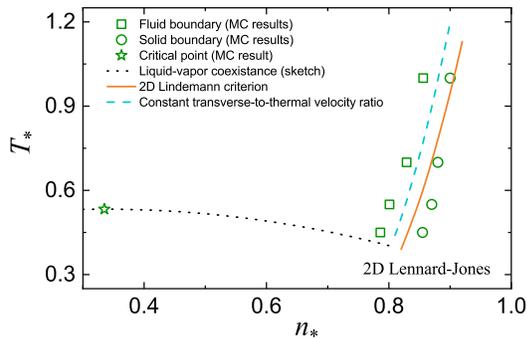}
\caption{Phase diagram of the 2D Lennard-Jones system in ($n_*$, $T_*$) plane. Symbols are the results from MC simulations~\cite{BarkerPhysA1981}. The solid curve corresponds to the 2D Lindemann's melting rule of Eqs.~(\ref{Lindemann1}) and (\ref{wD2}). The dashed line represents Eq.~(\ref{melting_tr}). The dotted curve shows approximate location of the liquid-vapor coexistance boundary. The critical point is located at $n_*\simeq 0.335$ and $T_*\simeq 0.533$; the triple point temperature is $T_{\rm tr}\simeq 0.415$~\cite{BarkerPhysA1981}. }
\label{Fig3}
\end{figure}

The applicability of the generalized Lindemann's melting criterion is not limited to systems with soft repulsive interactions. To demonstrate this we consider the conventional Lennard-Jones (LJ) potential, $\varphi(r)=4\epsilon\left[(\sigma/r)^{12}-(\sigma/r)^6\right]$, where $\epsilon$ and $\sigma$ are the energy and length scales, respectively. The sound velocities of a LJ solid can be expressed as $c_{l/t}^2/v_{\rm T}^2=({\mathcal A}_{l/t}n_*^{12/{\mathcal D}}-{\mathcal B}_{l/t}n_*^{6/{\mathcal D}})/T_*$, where the conventional reduced units $n_*=n\sigma^{\mathcal D}$ and $T_*=T/\epsilon$ are used. The constants ${\mathcal A}_{l/t}$ and ${\mathcal B}_{l/t}$ are expressed in terms of the corresponding lattice sums for $r^{-12}$ and $r^{-6}$ potentials. In 3D the constant transverse-to-thermal velocity ratio then implies freezing and melting equations of the form $T_*^{\rm L,S}= C_{12}^{\rm L,S}n_*^4-C_{6}^{\rm L,S}n_*^2$ (superscripts L and S correspond to liquid and solid, respectively ). This shape of the fluid-solid coexistence in 3D LJ systems with constant (or very weakly $n_*$-dependent) constants $C_{12}$ and $C_6$ is a very robust result reproduced in a number of various theories and approximations~\cite{PedersenNatCom2016,RosenfeldMolPhys1976,KhrapakJCP2011_2,HeyesPSS2015,KhrapakAIPAdv2016,CostigliolaPCCP2016}.       
Similarly, in 2D the freezing and melting equations are  $T_*^{\rm L,S}= C_{12}^{\rm L,S}n_*^6-C_{6}^{\rm L,S}n_*^3$. The melting curve, calculated from this expression using the {\it same} ratio $c_t/v_{\rm T}$ as for 2D Yukawa systems is plotted in the phase diagram of 2D LJ system in Fig.~\ref{Fig3}. The data shown correspond to the Monte Carlo (MC) calculation from Ref.~\cite{BarkerPhysA1981}. The curve falls into the fluid-solid coexistence region. If condition (\ref{melt2D}) is used instead, the theoretical curve moves much closer to the MC data related to the solid coexistence boundary. Overall, the agreement looks rather convincing.       
 
The only criterion known to date, which is applicable to the fluid-solid phase transition simultaneously in 3D and 2D, is the dynamical freezing criterion~\cite{LowenPRL1993,LowenPRE1996}. It states that the ratio of the long-time and short-time self-diffusion coefficients is about $0.1$ at freezing. This criterion is, however, only applicable to the overdamped systems exhibiting Brownian dynamics. In this sense the 2D Lindemann's melting criterion is more general, because it should apply to arbitrary level of frictional dissipation.     

An advantage of formulating Lindemann's law in terms of statistical mechanics is that a direct link to the thermodynamic properties is provided~\cite{RossPR1969}.  The Lindemann's law can be formulated in various ways, for instance as a quasi-universality of the reduced free volume, thermal component of the excess free energy, or excess entropy. In fact, using free-volume arguments,  one can immediately see that what appears under the logarithm of Eq.~(\ref{F2D}) is effectively the reduced MSD of a test particle from the {\it center of the cell formed by its neighboring particles}. This justifies previous heuristic approaches to the 2D Lindemann's melting rule~\cite{LozovikSSC1985,BedanovPLA1985,ZhengEPL1998}.  Quite importantly, however, present approach results in an explicit expression for the melting temperature $T_{\rm m}$.       


To conclude, the Lindemann's melting rule can be applied to the melting of 2D solids, when generalized in terms of statistical mechanics arguments. It produces an expression for the melting temperature, which formally coincides with that in 3D case and, thus, it belongs to very few melting indicators operating simultaneously in 3D and 2D.  The generalized 2D Lindemann's melting condition appears essentially equivalent to the BKTHNY condition of dislocation unbinding. A simple consequence of the Lindemann's melting condition is that the ratio of the transverse sound to the thermal velocity is approximately constant at the melting temperature in both 2D and 3D. This can be particularly useful in approximately locating the melting lines of various classical systems without performing accurate free energy calculations.  


\begin{acknowledgments} 
I would like to thank Boris Klumov for careful reading of the manuscript.
\end{acknowledgments}

\bibliographystyle{aipnum4-1}
\bibliography{Lindemann}

\providecommand{\noopsort}[1]{}\providecommand{\singleletter}[1]{#1}%
\begin{thebibliography}{78}%
\makeatletter
\providecommand \@ifxundefined [1]{%
 \@ifx{#1\undefined}
}%
\providecommand \@ifnum [1]{%
 \ifnum #1\expandafter \@firstoftwo
 \else \expandafter \@secondoftwo
 \fi
}%
\providecommand \@ifx [1]{%
 \ifx #1\expandafter \@firstoftwo
 \else \expandafter \@secondoftwo
 \fi
}%
\providecommand \natexlab [1]{#1}%
\providecommand \enquote  [1]{``#1''}%
\providecommand \bibnamefont  [1]{#1}%
\providecommand \bibfnamefont [1]{#1}%
\providecommand \citenamefont [1]{#1}%
\providecommand \href@noop [0]{\@secondoftwo}%
\providecommand \href [0]{\begingroup \@sanitize@url \@href}%
\providecommand \@href[1]{\@@startlink{#1}\@@href}%
\providecommand \@@href[1]{\endgroup#1\@@endlink}%
\providecommand \@sanitize@url [0]{\catcode `\\12\catcode `\$12\catcode
  `\&12\catcode `\#12\catcode `\^12\catcode `\_12\catcode `\%12\relax}%
\providecommand \@@startlink[1]{}%
\providecommand \@@endlink[0]{}%
\providecommand \url  [0]{\begingroup\@sanitize@url \@url }%
\providecommand \@url [1]{\endgroup\@href {#1}{\urlprefix }}%
\providecommand \urlprefix  [0]{URL }%
\providecommand \Eprint [0]{\href }%
\providecommand \doibase [0]{http://dx.doi.org/}%
\providecommand \selectlanguage [0]{\@gobble}%
\providecommand \bibinfo  [0]{\@secondoftwo}%
\providecommand \bibfield  [0]{\@secondoftwo}%
\providecommand \translation [1]{[#1]}%
\providecommand \BibitemOpen [0]{}%
\providecommand \bibitemStop [0]{}%
\providecommand \bibitemNoStop [0]{.\EOS\space}%
\providecommand \EOS [0]{\spacefactor3000\relax}%
\providecommand \BibitemShut  [1]{\csname bibitem#1\endcsname}%
\let\auto@bib@innerbib\@empty
\bibitem [{\citenamefont {Lindemann}(1910)}]{Lindemann}%
  \BibitemOpen
  \bibfield  {author} {\bibinfo {author} {\bibfnamefont {F.}~\bibnamefont
  {Lindemann}},\ }\href@noop {} {\bibfield  {journal} {\bibinfo  {journal} {Z.
  Phys.}\ }\textbf {\bibinfo {volume} {11}},\ \bibinfo {pages} {609} (\bibinfo
  {year} {1910})}\BibitemShut {NoStop}%
\bibitem [{\citenamefont {Peierls}(1934)}]{Peierls1934}%
  \BibitemOpen
  \bibfield  {author} {\bibinfo {author} {\bibfnamefont {R.}~\bibnamefont
  {Peierls}},\ }\href@noop {} {\bibfield  {journal} {\bibinfo  {journal} {Helv.
  Phys. Acta.}\ }\textbf {\bibinfo {volume} {7}},\ \bibinfo {pages} {81}
  (\bibinfo {year} {1934})}\BibitemShut {NoStop}%
\bibitem [{\citenamefont {Landau}(1937)}]{Landau1937}%
  \BibitemOpen
  \bibfield  {author} {\bibinfo {author} {\bibfnamefont {L.~D.}\ \bibnamefont
  {Landau}},\ }\href@noop {} {\bibfield  {journal} {\bibinfo  {journal} {J.
  Exp. Theor. Phys.}\ }\textbf {\bibinfo {volume} {7}},\ \bibinfo {pages} {627}
  (\bibinfo {year} {1937})}\BibitemShut {NoStop}%
\bibitem [{\citenamefont {Illing}\ \emph {et~al.}(2017)\citenamefont {Illing},
  \citenamefont {Fritschi}, \citenamefont {Kaiser}, \citenamefont {Klix},
  \citenamefont {Maret},\ and\ \citenamefont {Keim}}]{IllingPNAS2017}%
  \BibitemOpen
  \bibfield  {author} {\bibinfo {author} {\bibfnamefont {B.}~\bibnamefont
  {Illing}}, \bibinfo {author} {\bibfnamefont {S.}~\bibnamefont {Fritschi}},
  \bibinfo {author} {\bibfnamefont {H.}~\bibnamefont {Kaiser}}, \bibinfo
  {author} {\bibfnamefont {C.~L.}\ \bibnamefont {Klix}}, \bibinfo {author}
  {\bibfnamefont {G.}~\bibnamefont {Maret}}, \ and\ \bibinfo {author}
  {\bibfnamefont {P.}~\bibnamefont {Keim}},\ }\href {\doibase
  10.1073/pnas.1612964114} {\bibfield  {journal} {\bibinfo  {journal} {Proc.
  Natl. Acad. Sci. USA}\ }\textbf {\bibinfo {volume} {114}},\ \bibinfo {pages}
  {1856} (\bibinfo {year} {2017})}\BibitemShut {NoStop}%
\bibitem [{\citenamefont {Kosterlitz}\ and\ \citenamefont
  {Thouless}(1978)}]{Kosterlitz1978}%
  \BibitemOpen
  \bibfield  {author} {\bibinfo {author} {\bibfnamefont {J.}~\bibnamefont
  {Kosterlitz}}\ and\ \bibinfo {author} {\bibfnamefont {D.}~\bibnamefont
  {Thouless}},\ }\href {\doibase
  http://dx.doi.org/10.1016/S0079-6417(08)60175-4} {\bibfield  {journal}
  {\bibinfo  {journal} {Progress in Low Temperature Physics}\ }\textbf
  {\bibinfo {volume} {7}},\ \bibinfo {pages} {371 } (\bibinfo {year}
  {1978})}\BibitemShut {NoStop}%
\bibitem [{\citenamefont {Grimes}\ and\ \citenamefont
  {Adams}(1979)}]{GrimesPRL1979}%
  \BibitemOpen
  \bibfield  {author} {\bibinfo {author} {\bibfnamefont {C.~C.}\ \bibnamefont
  {Grimes}}\ and\ \bibinfo {author} {\bibfnamefont {G.}~\bibnamefont {Adams}},\
  }\href {\doibase 10.1103/physrevlett.42.795} {\bibfield  {journal} {\bibinfo
  {journal} {Phys. Rev. Lett.}\ }\textbf {\bibinfo {volume} {42}},\ \bibinfo
  {pages} {795} (\bibinfo {year} {1979})}\BibitemShut {NoStop}%
\bibitem [{\citenamefont {Zahn}, \citenamefont {Lenke},\ and\ \citenamefont
  {Maret}(1999)}]{ZahnPRL1999}%
  \BibitemOpen
  \bibfield  {author} {\bibinfo {author} {\bibfnamefont {K.}~\bibnamefont
  {Zahn}}, \bibinfo {author} {\bibfnamefont {R.}~\bibnamefont {Lenke}}, \ and\
  \bibinfo {author} {\bibfnamefont {G.}~\bibnamefont {Maret}},\ }\href
  {\doibase 10.1103/PhysRevLett.82.2721} {\bibfield  {journal} {\bibinfo
  {journal} {Phys. Rev. Lett.}\ }\textbf {\bibinfo {volume} {82}},\ \bibinfo
  {pages} {2721} (\bibinfo {year} {1999})}\BibitemShut {NoStop}%
\bibitem [{\citenamefont {Zahn}\ and\ \citenamefont
  {Maret}(2000)}]{ZahnPRL2000}%
  \BibitemOpen
  \bibfield  {author} {\bibinfo {author} {\bibfnamefont {K.}~\bibnamefont
  {Zahn}}\ and\ \bibinfo {author} {\bibfnamefont {G.}~\bibnamefont {Maret}},\
  }\href {\doibase 10.1103/PhysRevLett.85.3656} {\bibfield  {journal} {\bibinfo
   {journal} {Phys. Rev. Lett.}\ }\textbf {\bibinfo {volume} {85}},\ \bibinfo
  {pages} {3656} (\bibinfo {year} {2000})}\BibitemShut {NoStop}%
\bibitem [{\citenamefont {Kelleher}\ \emph {et~al.}(2017)\citenamefont
  {Kelleher}, \citenamefont {Guerra}, \citenamefont {Hollingsworth},\ and\
  \citenamefont {Chaikin}}]{KelleherPRE2017}%
  \BibitemOpen
  \bibfield  {author} {\bibinfo {author} {\bibfnamefont {C.~P.}\ \bibnamefont
  {Kelleher}}, \bibinfo {author} {\bibfnamefont {R.~E.}\ \bibnamefont
  {Guerra}}, \bibinfo {author} {\bibfnamefont {A.~D.}\ \bibnamefont
  {Hollingsworth}}, \ and\ \bibinfo {author} {\bibfnamefont {P.~M.}\
  \bibnamefont {Chaikin}},\ }\href {\doibase 10.1103/PhysRevE.95.022602}
  {\bibfield  {journal} {\bibinfo  {journal} {Phys. Rev. E}\ }\textbf {\bibinfo
  {volume} {95}},\ \bibinfo {pages} {022602} (\bibinfo {year}
  {2017})}\BibitemShut {NoStop}%
\bibitem [{\citenamefont {Chu}\ and\ \citenamefont {I}(1994)}]{ChuPRL1994}%
  \BibitemOpen
  \bibfield  {author} {\bibinfo {author} {\bibfnamefont {J.~H.}\ \bibnamefont
  {Chu}}\ and\ \bibinfo {author} {\bibfnamefont {L.}~\bibnamefont {I}},\ }\href
  {\doibase 10.1103/physrevlett.72.4009} {\bibfield  {journal} {\bibinfo
  {journal} {Phys. Rev. Lett.}\ }\textbf {\bibinfo {volume} {72}},\ \bibinfo
  {pages} {4009} (\bibinfo {year} {1994})}\BibitemShut {NoStop}%
\bibitem [{\citenamefont {Thomas}\ \emph {et~al.}(1994)\citenamefont {Thomas},
  \citenamefont {Morfill}, \citenamefont {Demmel}, \citenamefont {Goree},
  \citenamefont {Feuerbacher},\ and\ \citenamefont
  {M\"{o}hlmann}}]{ThomasPRL1994}%
  \BibitemOpen
  \bibfield  {author} {\bibinfo {author} {\bibfnamefont {H.}~\bibnamefont
  {Thomas}}, \bibinfo {author} {\bibfnamefont {G.~E.}\ \bibnamefont {Morfill}},
  \bibinfo {author} {\bibfnamefont {V.}~\bibnamefont {Demmel}}, \bibinfo
  {author} {\bibfnamefont {J.}~\bibnamefont {Goree}}, \bibinfo {author}
  {\bibfnamefont {B.}~\bibnamefont {Feuerbacher}}, \ and\ \bibinfo {author}
  {\bibfnamefont {D.}~\bibnamefont {M\"{o}hlmann}},\ }\href {\doibase
  10.1103/PhysRevLett.73.652} {\bibfield  {journal} {\bibinfo  {journal} {Phys.
  Rev. Lett.}\ }\textbf {\bibinfo {volume} {73}},\ \bibinfo {pages} {652}
  (\bibinfo {year} {1994})}\BibitemShut {NoStop}%
\bibitem [{\citenamefont {Melzer}, \citenamefont {Trottenberg},\ and\
  \citenamefont {Piel}(1994)}]{MelzerPLA1994}%
  \BibitemOpen
  \bibfield  {author} {\bibinfo {author} {\bibfnamefont {A.}~\bibnamefont
  {Melzer}}, \bibinfo {author} {\bibfnamefont {T.}~\bibnamefont {Trottenberg}},
  \ and\ \bibinfo {author} {\bibfnamefont {A.}~\bibnamefont {Piel}},\ }\href
  {\doibase 10.1016/0375-9601(94)90144-9} {\bibfield  {journal} {\bibinfo
  {journal} {Phys. Lett. A}\ }\textbf {\bibinfo {volume} {191}},\ \bibinfo
  {pages} {301} (\bibinfo {year} {1994})}\BibitemShut {NoStop}%
\bibitem [{\citenamefont {Hayashi}\ and\ \citenamefont
  {Tachibana}(1994)}]{Hayashi1994}%
  \BibitemOpen
  \bibfield  {author} {\bibinfo {author} {\bibfnamefont {Y.}~\bibnamefont
  {Hayashi}}\ and\ \bibinfo {author} {\bibfnamefont {K.}~\bibnamefont
  {Tachibana}},\ }\href {\doibase 10.1143/JJAP.33.L804} {\bibfield  {journal}
  {\bibinfo  {journal} {Jpn. J. Appl. Phys.}\ }\textbf {\bibinfo {volume}
  {33}},\ \bibinfo {pages} {L804} (\bibinfo {year} {1994})}\BibitemShut
  {NoStop}%
\bibitem [{\citenamefont {Thomas}\ and\ \citenamefont
  {Morfill}(1996)}]{ThomasNature1996}%
  \BibitemOpen
  \bibfield  {author} {\bibinfo {author} {\bibfnamefont {H.~M.}\ \bibnamefont
  {Thomas}}\ and\ \bibinfo {author} {\bibfnamefont {G.~E.}\ \bibnamefont
  {Morfill}},\ }\href {\doibase 10.1038/379806a0} {\bibfield  {journal}
  {\bibinfo  {journal} {Nature}\ }\textbf {\bibinfo {volume} {379}},\ \bibinfo
  {pages} {806} (\bibinfo {year} {1996})}\BibitemShut {NoStop}%
\bibitem [{\citenamefont {Fortov}\ \emph {et~al.}(2005)\citenamefont {Fortov},
  \citenamefont {Ivlev}, \citenamefont {Khrapak}, \citenamefont {Khrapak},\
  and\ \citenamefont {Morfill}}]{FortovPR}%
  \BibitemOpen
  \bibfield  {author} {\bibinfo {author} {\bibfnamefont {V.~E.}\ \bibnamefont
  {Fortov}}, \bibinfo {author} {\bibfnamefont {A.}~\bibnamefont {Ivlev}},
  \bibinfo {author} {\bibfnamefont {S.}~\bibnamefont {Khrapak}}, \bibinfo
  {author} {\bibfnamefont {A.}~\bibnamefont {Khrapak}}, \ and\ \bibinfo
  {author} {\bibfnamefont {G.}~\bibnamefont {Morfill}},\ }\href@noop {}
  {\bibfield  {journal} {\bibinfo  {journal} {Phys. Rep.}\ }\textbf {\bibinfo
  {volume} {421}},\ \bibinfo {pages} {1} (\bibinfo {year} {2005})}\BibitemShut
  {NoStop}%
\bibitem [{\citenamefont {Morfill}\ and\ \citenamefont
  {Ivlev}(2009)}]{MorfillRMP2009}%
  \BibitemOpen
  \bibfield  {author} {\bibinfo {author} {\bibfnamefont {G.~E.}\ \bibnamefont
  {Morfill}}\ and\ \bibinfo {author} {\bibfnamefont {A.~V.}\ \bibnamefont
  {Ivlev}},\ }\href {\doibase 10.1103/revmodphys.81.1353} {\bibfield  {journal}
  {\bibinfo  {journal} {Rev. Mod. Phys.}\ }\textbf {\bibinfo {volume} {81}},\
  \bibinfo {pages} {1353} (\bibinfo {year} {2009})}\BibitemShut {NoStop}%
\bibitem [{\citenamefont {Kosterlitz}(2017)}]{KosterlitzRMP2017}%
  \BibitemOpen
  \bibfield  {author} {\bibinfo {author} {\bibfnamefont {J.~M.}\ \bibnamefont
  {Kosterlitz}},\ }\href {\doibase 10.1103/revmodphys.89.040501} {\bibfield
  {journal} {\bibinfo  {journal} {Rev. Mod. Phys.}\ }\textbf {\bibinfo {volume}
  {89}},\ \bibinfo {pages} {040501} (\bibinfo {year} {2017})}\BibitemShut
  {NoStop}%
\bibitem [{\citenamefont {Berezinskii}(1971)}]{BerezinskiiJETP1971}%
  \BibitemOpen
  \bibfield  {author} {\bibinfo {author} {\bibfnamefont {V.~L.}\ \bibnamefont
  {Berezinskii}},\ }\href@noop {} {\bibfield  {journal} {\bibinfo  {journal}
  {J. Exp. Theor. Phys.}\ }\textbf {\bibinfo {volume} {32}},\ \bibinfo {pages}
  {493} (\bibinfo {year} {1971})}\BibitemShut {NoStop}%
\bibitem [{\citenamefont {Kosterlitz}\ and\ \citenamefont
  {Thouless}(1973)}]{KosterlitzJPC1973}%
  \BibitemOpen
  \bibfield  {author} {\bibinfo {author} {\bibfnamefont {J.~M.}\ \bibnamefont
  {Kosterlitz}}\ and\ \bibinfo {author} {\bibfnamefont {D.~J.}\ \bibnamefont
  {Thouless}},\ }\href {\doibase 10.1088/0022-3719/6/7/010} {\bibfield
  {journal} {\bibinfo  {journal} {J. Phys. C}\ }\textbf {\bibinfo {volume}
  {6}},\ \bibinfo {pages} {1181} (\bibinfo {year} {1973})}\BibitemShut
  {NoStop}%
\bibitem [{\citenamefont {Nelson}\ and\ \citenamefont
  {Halperin}(1979)}]{NelsonPRB1979}%
  \BibitemOpen
  \bibfield  {author} {\bibinfo {author} {\bibfnamefont {D.~R.}\ \bibnamefont
  {Nelson}}\ and\ \bibinfo {author} {\bibfnamefont {B.~I.}\ \bibnamefont
  {Halperin}},\ }\href {\doibase 10.1103/physrevb.19.2457} {\bibfield
  {journal} {\bibinfo  {journal} {Phys. Rev. B}\ }\textbf {\bibinfo {volume}
  {19}},\ \bibinfo {pages} {2457} (\bibinfo {year} {1979})}\BibitemShut
  {NoStop}%
\bibitem [{\citenamefont {Young}(1979)}]{YoungPRB1979}%
  \BibitemOpen
  \bibfield  {author} {\bibinfo {author} {\bibfnamefont {A.~P.}\ \bibnamefont
  {Young}},\ }\href {\doibase 10.1103/physrevb.19.1855} {\bibfield  {journal}
  {\bibinfo  {journal} {Phys. Rev. B}\ }\textbf {\bibinfo {volume} {19}},\
  \bibinfo {pages} {1855} (\bibinfo {year} {1979})}\BibitemShut {NoStop}%
\bibitem [{\citenamefont {Ryzhov}\ \emph {et~al.}(2017)\citenamefont {Ryzhov},
  \citenamefont {Tareyeva}, \citenamefont {Fomin},\ and\ \citenamefont
  {Tsiok}}]{RyzhovUFN2017}%
  \BibitemOpen
  \bibfield  {author} {\bibinfo {author} {\bibfnamefont {V.~N.}\ \bibnamefont
  {Ryzhov}}, \bibinfo {author} {\bibfnamefont {E.~E.}\ \bibnamefont
  {Tareyeva}}, \bibinfo {author} {\bibfnamefont {Y.~D.}\ \bibnamefont {Fomin}},
  \ and\ \bibinfo {author} {\bibfnamefont {E.~N.}\ \bibnamefont {Tsiok}},\
  }\href {\doibase 10.3367/ufne.2017.06.038161} {\bibfield  {journal} {\bibinfo
   {journal} {Phys.-Usp.}\ }\textbf {\bibinfo {volume} {60}},\ \bibinfo {pages}
  {857} (\bibinfo {year} {2017})}\BibitemShut {NoStop}%
\bibitem [{\citenamefont {Lin}, \citenamefont {Zheng},\ and\ \citenamefont
  {Trimper}(2006)}]{LinPRE2006}%
  \BibitemOpen
  \bibfield  {author} {\bibinfo {author} {\bibfnamefont {S.~Z.}\ \bibnamefont
  {Lin}}, \bibinfo {author} {\bibfnamefont {B.}~\bibnamefont {Zheng}}, \ and\
  \bibinfo {author} {\bibfnamefont {S.}~\bibnamefont {Trimper}},\ }\href
  {\doibase 10.1103/PhysRevE.73.066106} {\bibfield  {journal} {\bibinfo
  {journal} {Phys. Rev. E}\ }\textbf {\bibinfo {volume} {73}},\ \bibinfo
  {pages} {066106} (\bibinfo {year} {2006})}\BibitemShut {NoStop}%
\bibitem [{\citenamefont {Deutschl\"ander}\ \emph {et~al.}(2014)\citenamefont
  {Deutschl\"ander}, \citenamefont {Puertas}, \citenamefont {Maret},\ and\
  \citenamefont {Keim}}]{DeutschlanderPRL2014}%
  \BibitemOpen
  \bibfield  {author} {\bibinfo {author} {\bibfnamefont {S.}~\bibnamefont
  {Deutschl\"ander}}, \bibinfo {author} {\bibfnamefont {A.~M.}\ \bibnamefont
  {Puertas}}, \bibinfo {author} {\bibfnamefont {G.}~\bibnamefont {Maret}}, \
  and\ \bibinfo {author} {\bibfnamefont {P.}~\bibnamefont {Keim}},\ }\href
  {\doibase 10.1103/PhysRevLett.113.127801} {\bibfield  {journal} {\bibinfo
  {journal} {Phys. Rev. Lett.}\ }\textbf {\bibinfo {volume} {113}},\ \bibinfo
  {pages} {127801} (\bibinfo {year} {2014})}\BibitemShut {NoStop}%
\bibitem [{\citenamefont {Kelleher}\ \emph {et~al.}(2015)\citenamefont
  {Kelleher}, \citenamefont {Wang}, \citenamefont {Guerrero-Garc\'{\i}a},
  \citenamefont {Hollingsworth}, \citenamefont {Guerra}, \citenamefont
  {Krishnatreya}, \citenamefont {Grier}, \citenamefont {Manoharan},\ and\
  \citenamefont {Chaikin}}]{KelleherPRE2015}%
  \BibitemOpen
  \bibfield  {author} {\bibinfo {author} {\bibfnamefont {C.~P.}\ \bibnamefont
  {Kelleher}}, \bibinfo {author} {\bibfnamefont {A.}~\bibnamefont {Wang}},
  \bibinfo {author} {\bibfnamefont {G.~I.}\ \bibnamefont
  {Guerrero-Garc\'{\i}a}}, \bibinfo {author} {\bibfnamefont {A.~D.}\
  \bibnamefont {Hollingsworth}}, \bibinfo {author} {\bibfnamefont {R.~E.}\
  \bibnamefont {Guerra}}, \bibinfo {author} {\bibfnamefont {B.~J.}\
  \bibnamefont {Krishnatreya}}, \bibinfo {author} {\bibfnamefont {D.~G.}\
  \bibnamefont {Grier}}, \bibinfo {author} {\bibfnamefont {V.~N.}\ \bibnamefont
  {Manoharan}}, \ and\ \bibinfo {author} {\bibfnamefont {P.~M.}\ \bibnamefont
  {Chaikin}},\ }\href {\doibase 10.1103/PhysRevE.92.062306} {\bibfield
  {journal} {\bibinfo  {journal} {Phys. Rev. E}\ }\textbf {\bibinfo {volume}
  {92}},\ \bibinfo {pages} {062306} (\bibinfo {year} {2015})}\BibitemShut
  {NoStop}%
\bibitem [{\citenamefont {Kapfer}\ and\ \citenamefont
  {Krauth}(2015)}]{KapferPRL2015}%
  \BibitemOpen
  \bibfield  {author} {\bibinfo {author} {\bibfnamefont {S.~C.}\ \bibnamefont
  {Kapfer}}\ and\ \bibinfo {author} {\bibfnamefont {W.}~\bibnamefont
  {Krauth}},\ }\href {\doibase 10.1103/physrevlett.114.035702} {\bibfield
  {journal} {\bibinfo  {journal} {Phys. Rev. Lett.}\ }\textbf {\bibinfo
  {volume} {114}},\ \bibinfo {pages} {035702} (\bibinfo {year}
  {2015})}\BibitemShut {NoStop}%
\bibitem [{\citenamefont {Bernard}\ and\ \citenamefont
  {Krauth}(2011)}]{BernardPRL2011}%
  \BibitemOpen
  \bibfield  {author} {\bibinfo {author} {\bibfnamefont {E.~P.}\ \bibnamefont
  {Bernard}}\ and\ \bibinfo {author} {\bibfnamefont {W.}~\bibnamefont
  {Krauth}},\ }\href {\doibase 10.1103/physrevlett.107.155704} {\bibfield
  {journal} {\bibinfo  {journal} {Phys. Rev. Lett.}\ }\textbf {\bibinfo
  {volume} {107}},\ \bibinfo {pages} {155704} (\bibinfo {year}
  {2011})}\BibitemShut {NoStop}%
\bibitem [{\citenamefont {Engel}\ \emph {et~al.}(2013)\citenamefont {Engel},
  \citenamefont {Anderson}, \citenamefont {Glotzer}, \citenamefont {Isobe},
  \citenamefont {Bernard},\ and\ \citenamefont {Krauth}}]{EngelPRE2013}%
  \BibitemOpen
  \bibfield  {author} {\bibinfo {author} {\bibfnamefont {M.}~\bibnamefont
  {Engel}}, \bibinfo {author} {\bibfnamefont {J.~A.}\ \bibnamefont {Anderson}},
  \bibinfo {author} {\bibfnamefont {S.~C.}\ \bibnamefont {Glotzer}}, \bibinfo
  {author} {\bibfnamefont {M.}~\bibnamefont {Isobe}}, \bibinfo {author}
  {\bibfnamefont {E.~P.}\ \bibnamefont {Bernard}}, \ and\ \bibinfo {author}
  {\bibfnamefont {W.}~\bibnamefont {Krauth}},\ }\href {\doibase
  10.1103/PhysRevE.87.042134} {\bibfield  {journal} {\bibinfo  {journal} {Phys.
  Rev. E}\ }\textbf {\bibinfo {volume} {87}},\ \bibinfo {pages} {042134}
  (\bibinfo {year} {2013})}\BibitemShut {NoStop}%
\bibitem [{\citenamefont {Thorneywork}\ \emph {et~al.}(2017)\citenamefont
  {Thorneywork}, \citenamefont {Abbott}, \citenamefont {Aarts},\ and\
  \citenamefont {Dullens}}]{ThorneyworkPRL2017}%
  \BibitemOpen
  \bibfield  {author} {\bibinfo {author} {\bibfnamefont {A.~L.}\ \bibnamefont
  {Thorneywork}}, \bibinfo {author} {\bibfnamefont {J.~L.}\ \bibnamefont
  {Abbott}}, \bibinfo {author} {\bibfnamefont {D.~G. A.~L.}\ \bibnamefont
  {Aarts}}, \ and\ \bibinfo {author} {\bibfnamefont {R.~P.~A.}\ \bibnamefont
  {Dullens}},\ }\href {\doibase 10.1103/physrevlett.118.158001} {\bibfield
  {journal} {\bibinfo  {journal} {Phys. Rev. Lett.}\ }\textbf {\bibinfo
  {volume} {118}},\ \bibinfo {pages} {158001} (\bibinfo {year}
  {2017})}\BibitemShut {NoStop}%
\bibitem [{\citenamefont {Shiba}\ \emph {et~al.}(2016)\citenamefont {Shiba},
  \citenamefont {Yamada}, \citenamefont {Kawasaki},\ and\ \citenamefont
  {Kim}}]{ShibaPRL2016}%
  \BibitemOpen
  \bibfield  {author} {\bibinfo {author} {\bibfnamefont {H.}~\bibnamefont
  {Shiba}}, \bibinfo {author} {\bibfnamefont {Y.}~\bibnamefont {Yamada}},
  \bibinfo {author} {\bibfnamefont {T.}~\bibnamefont {Kawasaki}}, \ and\
  \bibinfo {author} {\bibfnamefont {K.}~\bibnamefont {Kim}},\ }\href {\doibase
  10.1103/physrevlett.117.245701} {\bibfield  {journal} {\bibinfo  {journal}
  {Phys. Rev. Lett.}\ }\textbf {\bibinfo {volume} {117}},\ \bibinfo {pages}
  {245701} (\bibinfo {year} {2016})}\BibitemShut {NoStop}%
\bibitem [{\citenamefont {Jancovici}(1967)}]{JancoviciPRL1967}%
  \BibitemOpen
  \bibfield  {author} {\bibinfo {author} {\bibfnamefont {B.}~\bibnamefont
  {Jancovici}},\ }\href {\doibase 10.1103/physrevlett.19.20} {\bibfield
  {journal} {\bibinfo  {journal} {Phys. Rev. Lett.}\ }\textbf {\bibinfo
  {volume} {19}},\ \bibinfo {pages} {20} (\bibinfo {year} {1967})}\BibitemShut
  {NoStop}%
\bibitem [{\citenamefont {Lozovik}\ and\ \citenamefont
  {Farztdinov}(1985)}]{LozovikSSC1985}%
  \BibitemOpen
  \bibfield  {author} {\bibinfo {author} {\bibfnamefont {Y.}~\bibnamefont
  {Lozovik}}\ and\ \bibinfo {author} {\bibfnamefont {V.}~\bibnamefont
  {Farztdinov}},\ }\href {\doibase 10.1016/0038-1098(85)90596-4} {\bibfield
  {journal} {\bibinfo  {journal} {Solid State Commun.}\ }\textbf {\bibinfo
  {volume} {54}},\ \bibinfo {pages} {725} (\bibinfo {year} {1985})}\BibitemShut
  {NoStop}%
\bibitem [{\citenamefont {Bedanov}, \citenamefont {Gadiyak},\ and\
  \citenamefont {Lozovik}(1985)}]{BedanovPLA1985}%
  \BibitemOpen
  \bibfield  {author} {\bibinfo {author} {\bibfnamefont {V.}~\bibnamefont
  {Bedanov}}, \bibinfo {author} {\bibfnamefont {G.}~\bibnamefont {Gadiyak}}, \
  and\ \bibinfo {author} {\bibfnamefont {Y.}~\bibnamefont {Lozovik}},\ }\href
  {\doibase 10.1016/0375-9601(85)90617-6} {\bibfield  {journal} {\bibinfo
  {journal} {Phys. Lett. A}\ }\textbf {\bibinfo {volume} {109}},\ \bibinfo
  {pages} {289} (\bibinfo {year} {1985})}\BibitemShut {NoStop}%
\bibitem [{\citenamefont {Goldoni}\ and\ \citenamefont
  {Peeters}(1996)}]{GoldoniPRB1996}%
  \BibitemOpen
  \bibfield  {author} {\bibinfo {author} {\bibfnamefont {G.}~\bibnamefont
  {Goldoni}}\ and\ \bibinfo {author} {\bibfnamefont {F.~M.}\ \bibnamefont
  {Peeters}},\ }\href {\doibase 10.1103/physrevb.53.4591} {\bibfield  {journal}
  {\bibinfo  {journal} {Phys. Rev. B}\ }\textbf {\bibinfo {volume} {53}},\
  \bibinfo {pages} {4591} (\bibinfo {year} {1996})}\BibitemShut {NoStop}%
\bibitem [{\citenamefont {Zheng}\ and\ \citenamefont
  {Earnshaw}(1998)}]{ZhengEPL1998}%
  \BibitemOpen
  \bibfield  {author} {\bibinfo {author} {\bibfnamefont {X.~H.}\ \bibnamefont
  {Zheng}}\ and\ \bibinfo {author} {\bibfnamefont {J.~C.}\ \bibnamefont
  {Earnshaw}},\ }\href {\doibase 10.1209/epl/i1998-00205-7} {\bibfield
  {journal} {\bibinfo  {journal} {Europhys. Lett. ({EPL})}\ }\textbf {\bibinfo
  {volume} {41}},\ \bibinfo {pages} {635} (\bibinfo {year} {1998})}\BibitemShut
  {NoStop}%
\bibitem [{\citenamefont {Ross}(1969)}]{RossPR1969}%
  \BibitemOpen
  \bibfield  {author} {\bibinfo {author} {\bibfnamefont {M.}~\bibnamefont
  {Ross}},\ }\href {\doibase 10.1103/physrev.184.233} {\bibfield  {journal}
  {\bibinfo  {journal} {Phys. Rev.}\ }\textbf {\bibinfo {volume} {184}},\
  \bibinfo {pages} {233} (\bibinfo {year} {1969})}\BibitemShut {NoStop}%
\bibitem [{\citenamefont {Young}\ and\ \citenamefont
  {Alder}(1974)}]{YoungJCP1974}%
  \BibitemOpen
  \bibfield  {author} {\bibinfo {author} {\bibfnamefont {D.~A.}\ \bibnamefont
  {Young}}\ and\ \bibinfo {author} {\bibfnamefont {B.~J.}\ \bibnamefont
  {Alder}},\ }\href {\doibase 10.1063/1.1681190} {\bibfield  {journal}
  {\bibinfo  {journal} {J. Chem. Phys.}\ }\textbf {\bibinfo {volume} {60}},\
  \bibinfo {pages} {1254} (\bibinfo {year} {1974})}\BibitemShut {NoStop}%
\bibitem [{\citenamefont {Buchenau}, \citenamefont {Zorn},\ and\ \citenamefont
  {Ramos}(2014)}]{BuchenauPRE2014}%
  \BibitemOpen
  \bibfield  {author} {\bibinfo {author} {\bibfnamefont {U.}~\bibnamefont
  {Buchenau}}, \bibinfo {author} {\bibfnamefont {R.}~\bibnamefont {Zorn}}, \
  and\ \bibinfo {author} {\bibfnamefont {M.~A.}\ \bibnamefont {Ramos}},\ }\href
  {\doibase 10.1103/physreve.90.042312} {\bibfield  {journal} {\bibinfo
  {journal} {Phys. Rev. E}\ }\textbf {\bibinfo {volume} {90}},\ \bibinfo
  {pages} {042312} (\bibinfo {year} {2014})}\BibitemShut {NoStop}%
\bibitem [{\citenamefont {Gann}, \citenamefont {Chakravarty},\ and\
  \citenamefont {Chester}(1979)}]{GannPRB1979}%
  \BibitemOpen
  \bibfield  {author} {\bibinfo {author} {\bibfnamefont {R.~C.}\ \bibnamefont
  {Gann}}, \bibinfo {author} {\bibfnamefont {S.}~\bibnamefont {Chakravarty}}, \
  and\ \bibinfo {author} {\bibfnamefont {G.~V.}\ \bibnamefont {Chester}},\
  }\href {\doibase 10.1103/physrevb.20.326} {\bibfield  {journal} {\bibinfo
  {journal} {Phys. Rev. B}\ }\textbf {\bibinfo {volume} {20}},\ \bibinfo
  {pages} {326} (\bibinfo {year} {1979})}\BibitemShut {NoStop}%
\bibitem [{\citenamefont {Toxvaerd}(1983)}]{ToxvaerdPRL1983}%
  \BibitemOpen
  \bibfield  {author} {\bibinfo {author} {\bibfnamefont {S.}~\bibnamefont
  {Toxvaerd}},\ }\href {\doibase 10.1103/physrevlett.51.1971} {\bibfield
  {journal} {\bibinfo  {journal} {Phys. Rev. Lett.}\ }\textbf {\bibinfo
  {volume} {51}},\ \bibinfo {pages} {1971} (\bibinfo {year}
  {1983})}\BibitemShut {NoStop}%
\bibitem [{\citenamefont {Dyre}(2014)}]{DyreJPCB2014}%
  \BibitemOpen
  \bibfield  {author} {\bibinfo {author} {\bibfnamefont {J.~C.}\ \bibnamefont
  {Dyre}},\ }\href {\doibase 10.1021/jp501852b} {\bibfield  {journal} {\bibinfo
   {journal} {J. Phys. Chem. B}\ }\textbf {\bibinfo {volume} {118}},\ \bibinfo
  {pages} {10007} (\bibinfo {year} {2014})}\BibitemShut {NoStop}%
\bibitem [{\citenamefont {Gnan}\ \emph {et~al.}(2009)\citenamefont {Gnan},
  \citenamefont {Schr{\o}der}, \citenamefont {Pedersen}, \citenamefont
  {Bailey},\ and\ \citenamefont {Dyre}}]{GnanJCP2009}%
  \BibitemOpen
  \bibfield  {author} {\bibinfo {author} {\bibfnamefont {N.}~\bibnamefont
  {Gnan}}, \bibinfo {author} {\bibfnamefont {T.~B.}\ \bibnamefont
  {Schr{\o}der}}, \bibinfo {author} {\bibfnamefont {U.~R.}\ \bibnamefont
  {Pedersen}}, \bibinfo {author} {\bibfnamefont {N.~P.}\ \bibnamefont
  {Bailey}}, \ and\ \bibinfo {author} {\bibfnamefont {J.~C.}\ \bibnamefont
  {Dyre}},\ }\href {\doibase 10.1063/1.3265957} {\bibfield  {journal} {\bibinfo
   {journal} {J. Chem. Phys.}\ }\textbf {\bibinfo {volume} {131}},\ \bibinfo
  {pages} {234504} (\bibinfo {year} {2009})}\BibitemShut {NoStop}%
\bibitem [{\citenamefont {Pedersen}\ \emph {et~al.}(2016)\citenamefont
  {Pedersen}, \citenamefont {Costigliola}, \citenamefont {Bailey},
  \citenamefont {Schr{\o}der},\ and\ \citenamefont
  {Dyre}}]{PedersenNatCom2016}%
  \BibitemOpen
  \bibfield  {author} {\bibinfo {author} {\bibfnamefont {U.~R.}\ \bibnamefont
  {Pedersen}}, \bibinfo {author} {\bibfnamefont {L.}~\bibnamefont
  {Costigliola}}, \bibinfo {author} {\bibfnamefont {N.~P.}\ \bibnamefont
  {Bailey}}, \bibinfo {author} {\bibfnamefont {T.~B.}\ \bibnamefont
  {Schr{\o}der}}, \ and\ \bibinfo {author} {\bibfnamefont {J.~C.}\ \bibnamefont
  {Dyre}},\ }\href {\doibase 10.1038/ncomms12386} {\bibfield  {journal}
  {\bibinfo  {journal} {Nature Commun.}\ }\textbf {\bibinfo {volume} {7}},\
  \bibinfo {pages} {12386} (\bibinfo {year} {2016})}\BibitemShut {NoStop}%
\bibitem [{\citenamefont {Landau}\ and\ \citenamefont
  {Lifshitz}(1986{\natexlab{a}})}]{LL_StatPhys}%
  \BibitemOpen
  \bibfield  {author} {\bibinfo {author} {\bibfnamefont {L.~D.}\ \bibnamefont
  {Landau}}\ and\ \bibinfo {author} {\bibfnamefont {E.}~\bibnamefont
  {Lifshitz}},\ }\href@noop {} {\emph {\bibinfo {title} {Statystical
  Physics}}}\ (\bibinfo  {publisher} {Amsterdam: Elsevier},\ \bibinfo {year}
  {1986})\BibitemShut {NoStop}%
\bibitem [{\citenamefont {Khrapak}, \citenamefont {Klumov},\ and\ \citenamefont
  {Couedel}(2017)}]{KhrapakSciRep2017}%
  \BibitemOpen
  \bibfield  {author} {\bibinfo {author} {\bibfnamefont {S.}~\bibnamefont
  {Khrapak}}, \bibinfo {author} {\bibfnamefont {B.}~\bibnamefont {Klumov}}, \
  and\ \bibinfo {author} {\bibfnamefont {L.}~\bibnamefont {Couedel}},\
  }\href@noop {} {\bibfield  {journal} {\bibinfo  {journal} {Sci. Reports}\
  }\textbf {\bibinfo {volume} {7}},\ \bibinfo {pages} {7985} (\bibinfo {year}
  {2017})}\BibitemShut {NoStop}%
\bibitem [{\citenamefont {Khrapak}(2019{\natexlab{a}})}]{KhrapakMolPhys2019}%
  \BibitemOpen
  \bibfield  {author} {\bibinfo {author} {\bibfnamefont {S.}~\bibnamefont
  {Khrapak}},\ }\href {\doibase 10.1080/00268976.2019.1643045} {\bibfield
  {journal} {\bibinfo  {journal} {Mol. Phys.}\ }\textbf {\bibinfo {volume}
  {xx}},\ \bibinfo {pages} {xxxx} (\bibinfo {year}
  {2019}{\natexlab{a}})}\BibitemShut {NoStop}%
\bibitem [{\citenamefont {Khrapak}(2019{\natexlab{b}})}]{KhrapakPoP2019}%
  \BibitemOpen
  \bibfield  {author} {\bibinfo {author} {\bibfnamefont {S.~A.}\ \bibnamefont
  {Khrapak}},\ }\href {\doibase 10.1063/1.5124676} {\bibfield  {journal}
  {\bibinfo  {journal} {Phys. Plasmas}\ }\textbf {\bibinfo {volume} {26}},\
  \bibinfo {pages} {103703} (\bibinfo {year} {2019}{\natexlab{b}})}\BibitemShut
  {NoStop}%
\bibitem [{\citenamefont {Peeters}\ and\ \citenamefont
  {Wu}(1987)}]{PeetersPRA1987}%
  \BibitemOpen
  \bibfield  {author} {\bibinfo {author} {\bibfnamefont {F.~M.}\ \bibnamefont
  {Peeters}}\ and\ \bibinfo {author} {\bibfnamefont {X.}~\bibnamefont {Wu}},\
  }\href {\doibase 10.1103/physreva.35.3109} {\bibfield  {journal} {\bibinfo
  {journal} {Phys. Rev. A}\ }\textbf {\bibinfo {volume} {35}},\ \bibinfo
  {pages} {3109} (\bibinfo {year} {1987})}\BibitemShut {NoStop}%
\bibitem [{\citenamefont {Landau}\ and\ \citenamefont
  {Lifshitz}(1986{\natexlab{b}})}]{LL_Elasticity}%
  \BibitemOpen
  \bibfield  {author} {\bibinfo {author} {\bibfnamefont {L.~D.}\ \bibnamefont
  {Landau}}\ and\ \bibinfo {author} {\bibfnamefont {E.}~\bibnamefont
  {Lifshitz}},\ }\href@noop {} {\emph {\bibinfo {title} {Theory of
  Elasticity}}}\ (\bibinfo  {publisher} {Amsterdam: Elsevier},\ \bibinfo {year}
  {1986})\BibitemShut {NoStop}%
\bibitem [{\citenamefont {Morf}(1979)}]{MorfPRL1979}%
  \BibitemOpen
  \bibfield  {author} {\bibinfo {author} {\bibfnamefont {R.~H.}\ \bibnamefont
  {Morf}},\ }\href {\doibase 10.1103/physrevlett.43.931} {\bibfield  {journal}
  {\bibinfo  {journal} {Phys. Rev. Lett.}\ }\textbf {\bibinfo {volume} {43}},\
  \bibinfo {pages} {931} (\bibinfo {year} {1979})}\BibitemShut {NoStop}%
\bibitem [{\citenamefont {Zanghellini}, \citenamefont {Keim},\ and\
  \citenamefont {von Gr\"{u}nberg}(2005)}]{ZanghelliniJPCM2005}%
  \BibitemOpen
  \bibfield  {author} {\bibinfo {author} {\bibfnamefont {J.}~\bibnamefont
  {Zanghellini}}, \bibinfo {author} {\bibfnamefont {P.}~\bibnamefont {Keim}}, \
  and\ \bibinfo {author} {\bibfnamefont {H.~H.}\ \bibnamefont {von
  Gr\"{u}nberg}},\ }\href {\doibase 10.1088/0953-8984/17/45/051} {\bibfield
  {journal} {\bibinfo  {journal} {J. Phys.: Condens. Matter}\ }\textbf
  {\bibinfo {volume} {17}},\ \bibinfo {pages} {S3579} (\bibinfo {year}
  {2005})}\BibitemShut {NoStop}%
\bibitem [{\citenamefont {Thouless}(1978)}]{ThoulessJPC1978}%
  \BibitemOpen
  \bibfield  {author} {\bibinfo {author} {\bibfnamefont {D.~J.}\ \bibnamefont
  {Thouless}},\ }\href {\doibase 10.1088/0022-3719/11/6/001} {\bibfield
  {journal} {\bibinfo  {journal} {J. Phys. C: Solid State Phys.}\ }\textbf
  {\bibinfo {volume} {11}},\ \bibinfo {pages} {L189} (\bibinfo {year}
  {1978})}\BibitemShut {NoStop}%
\bibitem [{\citenamefont {Khrapak}\ and\ \citenamefont
  {Khrapak}(2016)}]{KhrapakCPP2016}%
  \BibitemOpen
  \bibfield  {author} {\bibinfo {author} {\bibfnamefont {S.~A.}\ \bibnamefont
  {Khrapak}}\ and\ \bibinfo {author} {\bibfnamefont {A.~G.}\ \bibnamefont
  {Khrapak}},\ }\href {\doibase 10.1002/ctpp.201500104} {\bibfield  {journal}
  {\bibinfo  {journal} {Contrib. Plasma Phys.}\ }\textbf {\bibinfo {volume}
  {56}},\ \bibinfo {pages} {270} (\bibinfo {year} {2016})}\BibitemShut
  {NoStop}%
\bibitem [{\citenamefont {Khrapak}(2018)}]{KhrapakJCP2018_1}%
  \BibitemOpen
  \bibfield  {author} {\bibinfo {author} {\bibfnamefont {S.}~\bibnamefont
  {Khrapak}},\ }\href {\doibase 10.1063/1.5027201} {\bibfield  {journal}
  {\bibinfo  {journal} {J. Chem. Phys.}\ }\textbf {\bibinfo {volume} {148}},\
  \bibinfo {pages} {146101} (\bibinfo {year} {2018})}\BibitemShut {NoStop}%
\bibitem [{\citenamefont {Baus}\ and\ \citenamefont {Hansen}(1980)}]{Baus1980}%
  \BibitemOpen
  \bibfield  {author} {\bibinfo {author} {\bibfnamefont {M.}~\bibnamefont
  {Baus}}\ and\ \bibinfo {author} {\bibfnamefont {J.~P.}\ \bibnamefont
  {Hansen}},\ }\href {\doibase 10.1016/0370-1573(80)90022-8} {\bibfield
  {journal} {\bibinfo  {journal} {Phys. Rep.}\ }\textbf {\bibinfo {volume}
  {59}},\ \bibinfo {pages} {1} (\bibinfo {year} {1980})}\BibitemShut {NoStop}%
\bibitem [{\citenamefont {Khrapak}\ \emph {et~al.}(2018)\citenamefont
  {Khrapak}, \citenamefont {Kryuchkov}, \citenamefont {Mistryukova},
  \citenamefont {Khrapak},\ and\ \citenamefont {Yurchenko}}]{KhrapakJCP2018}%
  \BibitemOpen
  \bibfield  {author} {\bibinfo {author} {\bibfnamefont {S.~A.}\ \bibnamefont
  {Khrapak}}, \bibinfo {author} {\bibfnamefont {N.~P.}\ \bibnamefont
  {Kryuchkov}}, \bibinfo {author} {\bibfnamefont {L.~A.}\ \bibnamefont
  {Mistryukova}}, \bibinfo {author} {\bibfnamefont {A.~G.}\ \bibnamefont
  {Khrapak}}, \ and\ \bibinfo {author} {\bibfnamefont {S.~O.}\ \bibnamefont
  {Yurchenko}},\ }\href {\doibase 10.1063/1.5050708} {\bibfield  {journal}
  {\bibinfo  {journal} {J. Chem. Phys.}\ }\textbf {\bibinfo {volume} {149}},\
  \bibinfo {pages} {134114} (\bibinfo {year} {2018})}\BibitemShut {NoStop}%
\bibitem [{\citenamefont {Dubin}\ and\ \citenamefont
  {O'Neil}(1999)}]{DubinRMP1999}%
  \BibitemOpen
  \bibfield  {author} {\bibinfo {author} {\bibfnamefont {D.~H.~E.}\
  \bibnamefont {Dubin}}\ and\ \bibinfo {author} {\bibfnamefont {T.~M.}\
  \bibnamefont {O'Neil}},\ }\href {\doibase 10.1103/revmodphys.71.87}
  {\bibfield  {journal} {\bibinfo  {journal} {Rev. Mod. Phys.}\ }\textbf
  {\bibinfo {volume} {71}},\ \bibinfo {pages} {87} (\bibinfo {year}
  {1999})}\BibitemShut {NoStop}%
\bibitem [{\citenamefont {Dyre}(2006)}]{DyreRMP2006}%
  \BibitemOpen
  \bibfield  {author} {\bibinfo {author} {\bibfnamefont {J.~C.}\ \bibnamefont
  {Dyre}},\ }\href {\doibase 10.1103/revmodphys.78.953} {\bibfield  {journal}
  {\bibinfo  {journal} {Rev. Mod. Phys.}\ }\textbf {\bibinfo {volume} {78}},\
  \bibinfo {pages} {953} (\bibinfo {year} {2006})}\BibitemShut {NoStop}%
\bibitem [{\citenamefont {Hartmann}\ \emph {et~al.}(2005)\citenamefont
  {Hartmann}, \citenamefont {Kalman}, \citenamefont {Donk{\'{o}}},\ and\
  \citenamefont {Kutasi}}]{HartmannPRE2005}%
  \BibitemOpen
  \bibfield  {author} {\bibinfo {author} {\bibfnamefont {P.}~\bibnamefont
  {Hartmann}}, \bibinfo {author} {\bibfnamefont {G.~J.}\ \bibnamefont
  {Kalman}}, \bibinfo {author} {\bibfnamefont {Z.}~\bibnamefont {Donk{\'{o}}}},
  \ and\ \bibinfo {author} {\bibfnamefont {K.}~\bibnamefont {Kutasi}},\ }\href
  {\doibase 10.1103/physreve.72.026409} {\bibfield  {journal} {\bibinfo
  {journal} {Phys. Rev. E}\ }\textbf {\bibinfo {volume} {72}},\ \bibinfo
  {pages} {026409} (\bibinfo {year} {2005})}\BibitemShut {NoStop}%
\bibitem [{\citenamefont {Ivlev}\ \emph {et~al.}(2012)\citenamefont {Ivlev},
  \citenamefont {L\"{o}wen}, \citenamefont {Morfill},\ and\ \citenamefont
  {Royall}}]{IvlevBook}%
  \BibitemOpen
  \bibfield  {author} {\bibinfo {author} {\bibfnamefont {A.}~\bibnamefont
  {Ivlev}}, \bibinfo {author} {\bibfnamefont {H.}~\bibnamefont {L\"{o}wen}},
  \bibinfo {author} {\bibfnamefont {G.}~\bibnamefont {Morfill}}, \ and\
  \bibinfo {author} {\bibfnamefont {C.~P.}\ \bibnamefont {Royall}},\
  }\href@noop {} {\emph {\bibinfo {title} {Complex Plasmas and Colloidal
  Dispersions: Particle-Resolved Studies of Classical Liquids and Solids}}}\
  (\bibinfo  {publisher} {World Scientific},\ \bibinfo {year}
  {2012})\BibitemShut {NoStop}%
\bibitem [{\citenamefont {L\"{o}wen}, \citenamefont {Palberg},\ and\
  \citenamefont {Simon}(1993)}]{LowenPRL1993}%
  \BibitemOpen
  \bibfield  {author} {\bibinfo {author} {\bibfnamefont {H.}~\bibnamefont
  {L\"{o}wen}}, \bibinfo {author} {\bibfnamefont {T.}~\bibnamefont {Palberg}},
  \ and\ \bibinfo {author} {\bibfnamefont {R.}~\bibnamefont {Simon}},\ }\href
  {\doibase 10.1103/physrevlett.70.1557} {\bibfield  {journal} {\bibinfo
  {journal} {Phys. Rev. Lett.}\ }\textbf {\bibinfo {volume} {70}},\ \bibinfo
  {pages} {1557} (\bibinfo {year} {1993})}\BibitemShut {NoStop}%
\bibitem [{\citenamefont {Chaudhuri}\ \emph {et~al.}(2011)\citenamefont
  {Chaudhuri}, \citenamefont {Ivlev}, \citenamefont {Khrapak}, \citenamefont
  {Thomas},\ and\ \citenamefont {Morfill}}]{ChaudhuriSM2011}%
  \BibitemOpen
  \bibfield  {author} {\bibinfo {author} {\bibfnamefont {M.}~\bibnamefont
  {Chaudhuri}}, \bibinfo {author} {\bibfnamefont {A.~V.}\ \bibnamefont
  {Ivlev}}, \bibinfo {author} {\bibfnamefont {S.~A.}\ \bibnamefont {Khrapak}},
  \bibinfo {author} {\bibfnamefont {H.~M.}\ \bibnamefont {Thomas}}, \ and\
  \bibinfo {author} {\bibfnamefont {G.~E.}\ \bibnamefont {Morfill}},\ }\href
  {\doibase 10.1039/c0sm00813c} {\bibfield  {journal} {\bibinfo  {journal}
  {Soft Matter}\ }\textbf {\bibinfo {volume} {7}},\ \bibinfo {pages} {1287}
  (\bibinfo {year} {2011})}\BibitemShut {NoStop}%
\bibitem [{\citenamefont {Robbins}, \citenamefont {Kremer},\ and\ \citenamefont
  {Grest}(1988)}]{RobbinsJCP1988}%
  \BibitemOpen
  \bibfield  {author} {\bibinfo {author} {\bibfnamefont {M.~O.}\ \bibnamefont
  {Robbins}}, \bibinfo {author} {\bibfnamefont {K.}~\bibnamefont {Kremer}}, \
  and\ \bibinfo {author} {\bibfnamefont {G.~S.}\ \bibnamefont {Grest}},\ }\href
  {\doibase 10.1063/1.453924} {\bibfield  {journal} {\bibinfo  {journal} {J.
  Chem. Phys.}\ }\textbf {\bibinfo {volume} {88}},\ \bibinfo {pages} {3286}
  (\bibinfo {year} {1988})}\BibitemShut {NoStop}%
\bibitem [{\citenamefont {Meijer}\ and\ \citenamefont
  {Frenkel}(1991)}]{MeijerJCP1991}%
  \BibitemOpen
  \bibfield  {author} {\bibinfo {author} {\bibfnamefont {E.~J.}\ \bibnamefont
  {Meijer}}\ and\ \bibinfo {author} {\bibfnamefont {D.}~\bibnamefont
  {Frenkel}},\ }\href {\doibase 10.1063/1.459898} {\bibfield  {journal}
  {\bibinfo  {journal} {J. Chem. Phys.}\ }\textbf {\bibinfo {volume} {94}},\
  \bibinfo {pages} {2269} (\bibinfo {year} {1991})}\BibitemShut {NoStop}%
\bibitem [{\citenamefont {Hamaguchi}, \citenamefont {Farouki},\ and\
  \citenamefont {Dubin}(1997)}]{HamaguchiPRE1997}%
  \BibitemOpen
  \bibfield  {author} {\bibinfo {author} {\bibfnamefont {S.}~\bibnamefont
  {Hamaguchi}}, \bibinfo {author} {\bibfnamefont {R.~T.}\ \bibnamefont
  {Farouki}}, \ and\ \bibinfo {author} {\bibfnamefont {D.~H.~E.}\ \bibnamefont
  {Dubin}},\ }\href {\doibase 10.1103/physreve.56.4671} {\bibfield  {journal}
  {\bibinfo  {journal} {Phys. Rev. E}\ }\textbf {\bibinfo {volume} {56}},\
  \bibinfo {pages} {4671} (\bibinfo {year} {1997})}\BibitemShut {NoStop}%
\bibitem [{\citenamefont {Vaulina}, \citenamefont {Khrapak},\ and\
  \citenamefont {Morfill}(2002)}]{VaulinaPRE2002}%
  \BibitemOpen
  \bibfield  {author} {\bibinfo {author} {\bibfnamefont {O.}~\bibnamefont
  {Vaulina}}, \bibinfo {author} {\bibfnamefont {S.}~\bibnamefont {Khrapak}}, \
  and\ \bibinfo {author} {\bibfnamefont {G.}~\bibnamefont {Morfill}},\ }\href
  {\doibase 10.1103/physreve.66.016404} {\bibfield  {journal} {\bibinfo
  {journal} {Phys. Rev. E}\ }\textbf {\bibinfo {volume} {66}},\ \bibinfo
  {pages} {016404} (\bibinfo {year} {2002})}\BibitemShut {NoStop}%
\bibitem [{\citenamefont {Khrapak}\ and\ \citenamefont
  {Morfill}(2009)}]{KhrapakPRL2009}%
  \BibitemOpen
  \bibfield  {author} {\bibinfo {author} {\bibfnamefont {S.~A.}\ \bibnamefont
  {Khrapak}}\ and\ \bibinfo {author} {\bibfnamefont {G.~E.}\ \bibnamefont
  {Morfill}},\ }\href {\doibase 10.1103/physrevlett.103.255003} {\bibfield
  {journal} {\bibinfo  {journal} {Phys. Rev. Lett.}\ }\textbf {\bibinfo
  {volume} {103}},\ \bibinfo {pages} {255003} (\bibinfo {year}
  {2009})}\BibitemShut {NoStop}%
\bibitem [{\citenamefont {Yazdi}\ \emph {et~al.}(2014)\citenamefont {Yazdi},
  \citenamefont {Ivlev}, \citenamefont {Khrapak}, \citenamefont {Thomas},
  \citenamefont {Morfill}, \citenamefont {L\"{o}wen}, \citenamefont {Wysocki},\
  and\ \citenamefont {Sperl}}]{YazdiPRE2014}%
  \BibitemOpen
  \bibfield  {author} {\bibinfo {author} {\bibfnamefont {A.}~\bibnamefont
  {Yazdi}}, \bibinfo {author} {\bibfnamefont {A.}~\bibnamefont {Ivlev}},
  \bibinfo {author} {\bibfnamefont {S.}~\bibnamefont {Khrapak}}, \bibinfo
  {author} {\bibfnamefont {H.}~\bibnamefont {Thomas}}, \bibinfo {author}
  {\bibfnamefont {G.~E.}\ \bibnamefont {Morfill}}, \bibinfo {author}
  {\bibfnamefont {H.}~\bibnamefont {L\"{o}wen}}, \bibinfo {author}
  {\bibfnamefont {A.}~\bibnamefont {Wysocki}}, \ and\ \bibinfo {author}
  {\bibfnamefont {M.}~\bibnamefont {Sperl}},\ }\href {\doibase
  10.1103/physreve.89.063105} {\bibfield  {journal} {\bibinfo  {journal} {Phys.
  Rev. E}\ }\textbf {\bibinfo {volume} {89}},\ \bibinfo {pages} {063105}
  (\bibinfo {year} {2014})}\BibitemShut {NoStop}%
\bibitem [{\citenamefont {Yurchenko}, \citenamefont {Kryuchkov},\ and\
  \citenamefont {Ivlev}(2016)}]{YurchenkoJPCM2016}%
  \BibitemOpen
  \bibfield  {author} {\bibinfo {author} {\bibfnamefont {S.~O.}\ \bibnamefont
  {Yurchenko}}, \bibinfo {author} {\bibfnamefont {N.~P.}\ \bibnamefont
  {Kryuchkov}}, \ and\ \bibinfo {author} {\bibfnamefont {A.~V.}\ \bibnamefont
  {Ivlev}},\ }\href {\doibase 10.1088/0953-8984/28/23/235401} {\bibfield
  {journal} {\bibinfo  {journal} {J. Phys.: Condens. Matter}\ }\textbf
  {\bibinfo {volume} {28}},\ \bibinfo {pages} {235401} (\bibinfo {year}
  {2016})}\BibitemShut {NoStop}%
\bibitem [{\citenamefont {Kryuchkov}, \citenamefont {Khrapak},\ and\
  \citenamefont {Yurchenko}(2017)}]{KryuchkovJCP2017}%
  \BibitemOpen
  \bibfield  {author} {\bibinfo {author} {\bibfnamefont {N.~P.}\ \bibnamefont
  {Kryuchkov}}, \bibinfo {author} {\bibfnamefont {S.~A.}\ \bibnamefont
  {Khrapak}}, \ and\ \bibinfo {author} {\bibfnamefont {S.~O.}\ \bibnamefont
  {Yurchenko}},\ }\href {\doibase 10.1063/1.4979325} {\bibfield  {journal}
  {\bibinfo  {journal} {J. Chem. Phys.}\ }\textbf {\bibinfo {volume} {146}},\
  \bibinfo {pages} {134702} (\bibinfo {year} {2017})}\BibitemShut {NoStop}%
\bibitem [{\citenamefont {Khrapak}\ and\ \citenamefont
  {Klumov}(2018)}]{KhrapakPoP2018}%
  \BibitemOpen
  \bibfield  {author} {\bibinfo {author} {\bibfnamefont {S.}~\bibnamefont
  {Khrapak}}\ and\ \bibinfo {author} {\bibfnamefont {B.}~\bibnamefont
  {Klumov}},\ }\href {\doibase 10.1063/1.5025396} {\bibfield  {journal}
  {\bibinfo  {journal} {Phys. Plasmas}\ }\textbf {\bibinfo {volume} {25}},\
  \bibinfo {pages} {033706} (\bibinfo {year} {2018})}\BibitemShut {NoStop}%
\bibitem [{\citenamefont {Rosenfeld}(1976)}]{RosenfeldMolPhys1976}%
  \BibitemOpen
  \bibfield  {author} {\bibinfo {author} {\bibfnamefont {Y.}~\bibnamefont
  {Rosenfeld}},\ }\href {\doibase 10.1080/00268977600102381} {\bibfield
  {journal} {\bibinfo  {journal} {Mol. Phys.}\ }\textbf {\bibinfo {volume}
  {32}},\ \bibinfo {pages} {963} (\bibinfo {year} {1976})}\BibitemShut
  {NoStop}%
\bibitem [{\citenamefont {Khrapak}\ and\ \citenamefont
  {Morfill}(2011)}]{KhrapakJCP2011_2}%
  \BibitemOpen
  \bibfield  {author} {\bibinfo {author} {\bibfnamefont {S.~A.}\ \bibnamefont
  {Khrapak}}\ and\ \bibinfo {author} {\bibfnamefont {G.~E.}\ \bibnamefont
  {Morfill}},\ }\href {\doibase 10.1063/1.3561698} {\bibfield  {journal}
  {\bibinfo  {journal} {J. Chem. Phys.}\ }\textbf {\bibinfo {volume} {134}},\
  \bibinfo {pages} {094108} (\bibinfo {year} {2011})}\BibitemShut {NoStop}%
\bibitem [{\citenamefont {Heyes}, \citenamefont {Dini},\ and\ \citenamefont
  {Bra{\'{n}}ka}(2015)}]{HeyesPSS2015}%
  \BibitemOpen
  \bibfield  {author} {\bibinfo {author} {\bibfnamefont {D.~M.}\ \bibnamefont
  {Heyes}}, \bibinfo {author} {\bibfnamefont {D.}~\bibnamefont {Dini}}, \ and\
  \bibinfo {author} {\bibfnamefont {A.~C.}\ \bibnamefont {Bra{\'{n}}ka}},\
  }\href {\doibase 10.1002/pssb.201451695} {\bibfield  {journal} {\bibinfo
  {journal} {Phys. Status Solidi (b)}\ }\textbf {\bibinfo {volume} {252}},\
  \bibinfo {pages} {1514} (\bibinfo {year} {2015})}\BibitemShut {NoStop}%
\bibitem [{\citenamefont {Khrapak}\ and\ \citenamefont
  {Ning}(2016)}]{KhrapakAIPAdv2016}%
  \BibitemOpen
  \bibfield  {author} {\bibinfo {author} {\bibfnamefont {S.~A.}\ \bibnamefont
  {Khrapak}}\ and\ \bibinfo {author} {\bibfnamefont {N.}~\bibnamefont {Ning}},\
  }\href {\doibase 10.1063/1.4952587} {\bibfield  {journal} {\bibinfo
  {journal} {{AIP} Adv.}\ }\textbf {\bibinfo {volume} {6}},\ \bibinfo {pages}
  {055215} (\bibinfo {year} {2016})}\BibitemShut {NoStop}%
\bibitem [{\citenamefont {Costigliola}, \citenamefont {Schr{\o}der},\ and\
  \citenamefont {Dyre}(2016)}]{CostigliolaPCCP2016}%
  \BibitemOpen
  \bibfield  {author} {\bibinfo {author} {\bibfnamefont {L.}~\bibnamefont
  {Costigliola}}, \bibinfo {author} {\bibfnamefont {T.~B.}\ \bibnamefont
  {Schr{\o}der}}, \ and\ \bibinfo {author} {\bibfnamefont {J.~C.}\ \bibnamefont
  {Dyre}},\ }\href {\doibase 10.1039/c5cp06363a} {\bibfield  {journal}
  {\bibinfo  {journal} {Phys. Chem. Chem. Phys.}\ }\textbf {\bibinfo {volume}
  {18}},\ \bibinfo {pages} {14678} (\bibinfo {year} {2016})}\BibitemShut
  {NoStop}%
\bibitem [{\citenamefont {Barker}, \citenamefont {Henderson},\ and\
  \citenamefont {Abraham}(1981)}]{BarkerPhysA1981}%
  \BibitemOpen
  \bibfield  {author} {\bibinfo {author} {\bibfnamefont {J.}~\bibnamefont
  {Barker}}, \bibinfo {author} {\bibfnamefont {D.}~\bibnamefont {Henderson}}, \
  and\ \bibinfo {author} {\bibfnamefont {F.}~\bibnamefont {Abraham}},\ }\href
  {\doibase 10.1016/0378-4371(81)90222-3} {\bibfield  {journal} {\bibinfo
  {journal} {Phys. A}\ }\textbf {\bibinfo {volume} {106}},\ \bibinfo {pages}
  {226} (\bibinfo {year} {1981})}\BibitemShut {NoStop}%
\bibitem [{\citenamefont {L\"{o}wen}(1996)}]{LowenPRE1996}%
  \BibitemOpen
  \bibfield  {author} {\bibinfo {author} {\bibfnamefont {H.}~\bibnamefont
  {L\"{o}wen}},\ }\href {\doibase 10.1103/physreve.53.r29} {\bibfield
  {journal} {\bibinfo  {journal} {Phys. Rev. E}\ }\textbf {\bibinfo {volume}
  {53}},\ \bibinfo {pages} {R29} (\bibinfo {year} {1996})}\BibitemShut
  {NoStop}%
\end{thebibliography}%

\end{document}